\documentclass[12pt]{article}

\usepackage{graphicx}

\usepackage[russian,english]{babel}

\begin{document}

\selectlanguage{english}

\begin{center} \bf
FROISSART BOUNDS \\ FOR AMPLITUDES AND CROSS SECTIONS\\
AT HIGH ENERGIES~\footnote{Based on the lecture presented
at the XLV Winter School of PNPI (March 2011) and on the
seminar talk at the Ruhr University Bochum (October 2011).
The  Russian version is published in ``Nuclear and particle
physics, Theoretical physics (Proceedings of the XLV
Winter School of PNPI, Feb. 28 - March 5, 2011)'', ed.
V.T.Kim, PNPI, St.Petersburg, 2011, pp.20-26.}

\bigskip
Ya. I. Azimov  (PNPI)

\bigskip
A b s t r a c t
\end{center}

High-energy behavior of total cross sections is
discussed in experiment and theory. Origin and
meaning of the Froissart bounds are described and
explained. Violation of the familiar log-squared
bound appears to not violate unitarity (contrary
to the common opinion), but correspond to rapid
high-energy increase of the amplitude in nonphysical
regions.

\bigskip

\newpage

The elementary particle physics (or, the same, high
energy physics) is considered as a separate branch
of physics since 1956, when the Rochester University,
USA, organized the Conference on High Energy Physics
(since then the ``Rochester'' Conferences gathered
once a year in different cities and countries; after
1964 they are biennials called ``International
Conferences on High Energy Physics''). But, sure,
elementary particles and their interactions had been
investigated even before 1956. It became clear as
early as in '30-ies that particles have interactions
of several different kinds. And it was discovered in
'40-ies that strong interactions with increasing energy
provide increasing multiple meson production. In other
words, the role of inelastic processes grows with
growing energy in collisions of strong-interacting
particles (they are called ``hadrons'' since 1962,
according to suggestion of L.B. Okun).

To 1960, the idea had been formed that scattering of
hadrons at very high energies should be similar to the
classical diffraction of light on a black (completely
absorbing) disc of a finite radius. If this were true,
the total interaction cross sections at very high energies
should be asymptotically constant, and the elastic cross
sections should be a fixed part of the total ones. Angular
distribution (or distribution in the momentum transfer
$-t$) should look as the diffraction peak $\sim\exp(bt)$
with a constant slope $b$, which is proportional to the
radius squared of the disc. Experimental data of those
years (in the then available energy interval) seemed to
agree with such expectations.

However, in 1961 there appeared two theoretical papers
which cast doubts on applicability of such a simple picture.
One of them was presented by \mbox{V.N. Gribov~\cite{BH}.}
It showed that the classical diffraction is incompatible
with the analytical properties of hadron amplitudes
when combined with the cross-channel unitarity condition.
This result has become an impetus to construct the Reggeon
theory, according to which the diffraction peak changes
(shrinks) with increasing energy, even if the total cross
section is asymptotically constant.

The other paper was published by Marcel Froissart~\cite{Fr}.
Froissart (he is, by the way, a member of the old noble
French family) began his work with the hypothesis that
total cross sections of hadron interactions may infinitely
grow with energy (though no experimental evidence for such
possibility had been seen to that time). Then he applied
the unitarity condition together with the analyticity of an
elastic amplitude, as expressed by the dispersion relations
with a finite number of subtractions. Based on such,
seemingly very ``soft'', conditions (nearly from nothing)
Froissart was able to receive quite tangible restriction
for a possible energy growth rate of the forward (backward)
scattering amplitude, and even stronger restrictions for the
fixed angle nonforward (nonbackward) scattering. Since the
unitarity condition (the optical theorem) relates the forward
elastic amplitude with the total interaction cross section,
it appeared that the total cross sections might not grow faster
than the logarithm squared of the energy. This result, known
as ``the Froissart theorem'', has become one of key points
when constructing theoretical models for high-energy strong
interactions. Moreover, it became a sincere belief for public
opinion of the high-energy physics community, that violation
of the Froissart theorem would mean violation of unitarity.

In the years after 1961, our knowledge of strong interactions
has been significantly expanded to higher energies. The following
experimental facts have been definitely established.

\begin{itemize}

\item The diffraction peaks shrink indeed with growing energy;
their slopes in respect to the momentum transfer grow at least
as the logarithm of energy.

\item The total cross sections, as is clear now, indeed increase
with energy. Existing data for different hadrons agree with
the hypothesis that the total cross sections asymptotically
grow as $\ln^2 s~$ ($s$ is the c.m.s. energy squared).

\end{itemize}

Most advanced in the energy scale are investigations of
nucleon-(anti)nucleon interactions, especially if one adds
data from cosmic ray studies. Existing values for the total
$pp$ and $p\bar{p}$ cross sections may be quite satisfactorily
described by the curves shown in Fig.1 (taken from
Ref.\cite{block}). Their high-energy behavior is proportional
to the log-squared energy.

However, the accelerator energy interval available in the
pre-LHC era is rather narrow in the logarithmic scale, while
data extracted from cosmic ray experiments have great
uncertainties. As a result, significant ambiguities may (and do)
appear in the description of the data. In particular, possible
are ``heretic'' descriptions, which contradict to the canonically
understood Froissart result. For example, Fig.2 (taken from
Ref.\cite{land}) shows such a description of the total cross
sections which corresponds to the power increase with energy as
$s^{\delta}$, though with a small exponent $\delta\approx0.08$.
Thus, experiments have not allowed yet to reach a definite
conclusion, whether the log-squared energy asymptotics is true
or not.

LHC extends the accelerator energies to the values which have
been available earlier, but only in cosmic rays. Meanwhile,
the accelerator measurements

\begin{figure}[h]
\centering
\includegraphics[width=0.70\hsize]{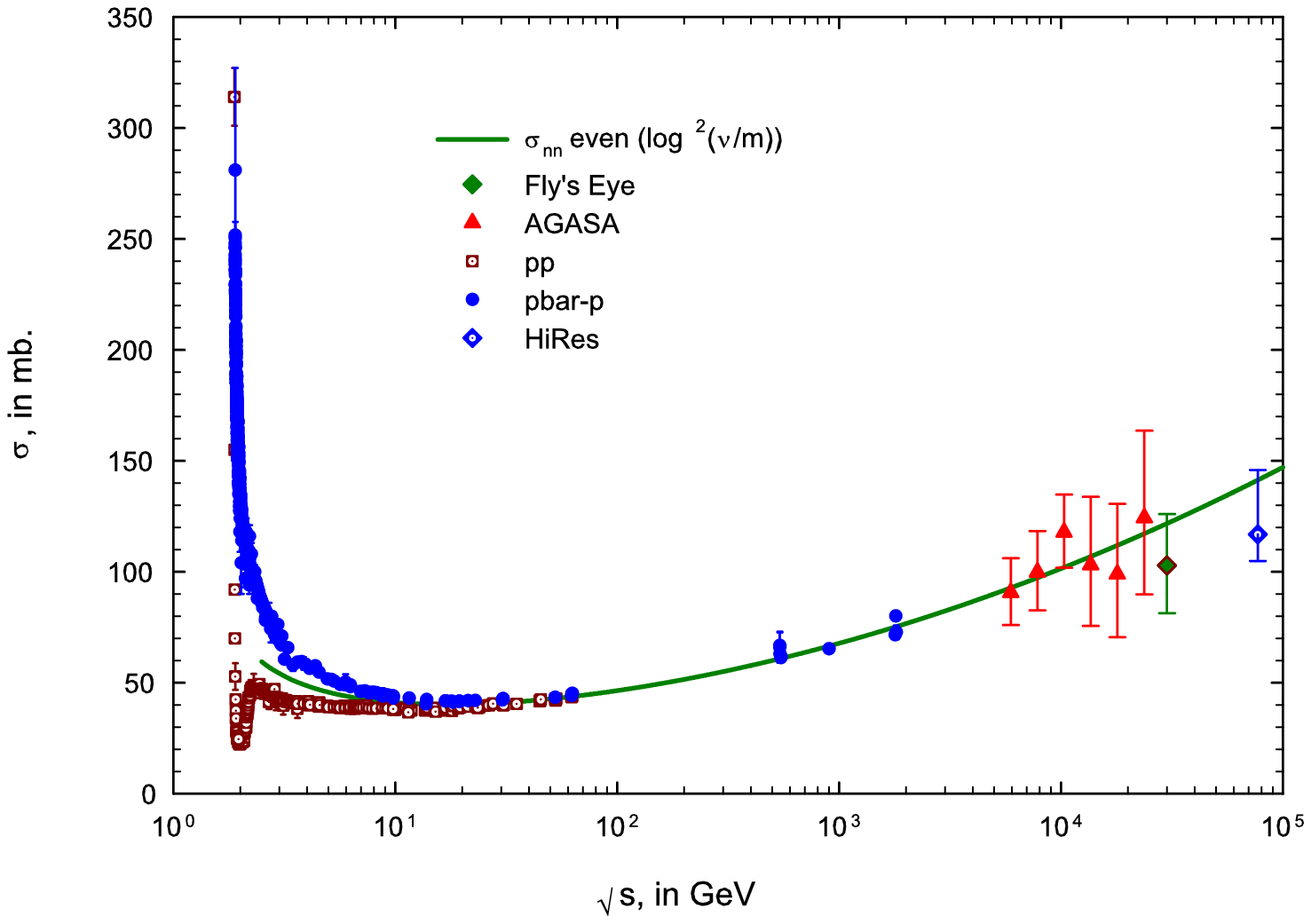}
\vspace*{-0.3cm}
\caption{\small  Fit for all data on the total $pp$
and $p\bar{p}$ cross sections available before
LHC~\cite{block}. The curves asymptotically grow
as $\ln^2 s$.}


\vspace*{1cm}
\centering
\includegraphics[width=0.70\hsize]{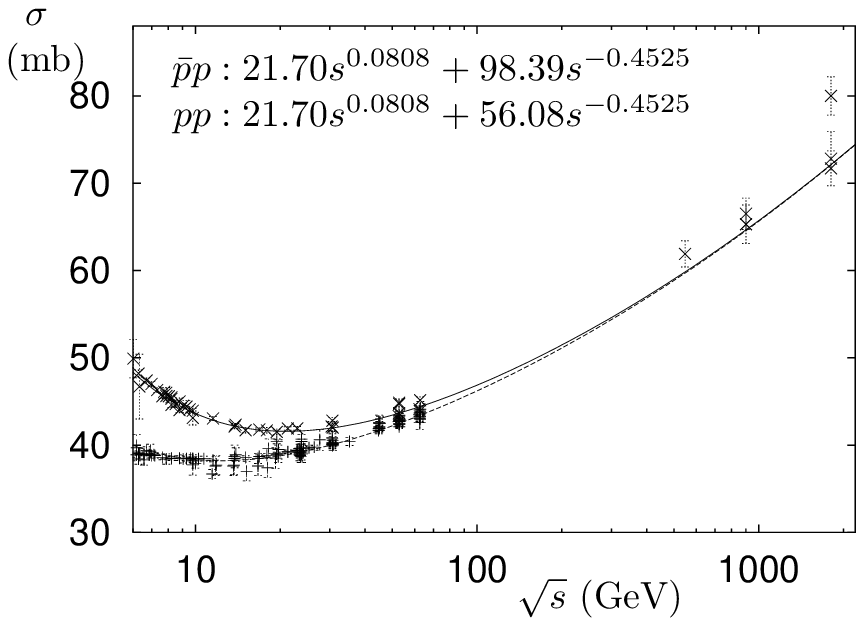}
\vspace*{-0.3cm}
\caption{\small  Fit for all acceleratot data
on the total $pp$ and $p\bar{p}$ cross sections
available before LHC~\cite{land}. The curves
asymptotically grow as a power of energy.}

\end{figure}
\clearpage
\noindent
are much more precise. Therefore, one can hope that
the LHC data, especially at its maximal energy (not
reached yet), may be able to clarify the situation.

It is interesting (and useful), however, to examine also the
theoretical basis of the Froissart theorem. This was just the
aim of the paper~\cite{az}, which revises derivation of the
theorem and meaning of its results. The paper may be easily
reached either in the journal or as the arXiv e-print, so it is
not necessary to present here all its calculations and formulas.
Instead, it is sufficient to describe the main results and
conclusions of the paper.

\begin{itemize}

\item  A necessary {\it physical} input for the Froissart
theorem is, of course, unitarity. It works in two ways: on one
side, the scattering-channel unitarity restricts elastic
partial-wave amplitudes; on the other side, the cross-channel
unitarity relates positions for scattering angle singularities
of the elastic amplitude with the mass spectrum in the cross
channel.

\item  Another {\it physical} input is the absence of massless
particles. It guaranties the absence of angle singularities
both inside the physical region and on its edges.

\item  A necessary {\it mathematical} base for the Froissart
theorem is provided by properties of the Legendre functions.
Especially important appears the behavior of $P_l(z)$ at
$l\to+\infty$. The infinite point in the $l$-plane is an
essential singularity for the Legendre functions. As a result,
their asymptotic forms at large positive $l$ are sharply
different in the three cases: inside the $z$-interval
$(-1, +1)$, at its edges ({\it i.e.}, at $z=\pm1$), and outside
this interval, though the points $z=\pm1$ are not singular for
$P_l(z)$ with physical (integer positive) values of $l$. On one
side, therefore, discontinuities become possible (and arise
indeed) between high-energy asymptotics of an elastic amplitude
in the three configurations: inside the physical region of
angles, at its boundary ({\it i.e.}, for the forward or backward
scattering), at nonphysical (complex) angles (it is worth to
emphasize that transitions between those three configurations
do not touch any singularities of the amplitude). On the other
side, due to properties of $P_l(z)$, the rate of high-energy
increase of the amplitude is much more moderate for physical
angles than for nonphysical ones. Such sharp moderateness of
the amplitudes in physical configurations is just {\it the
true meaning} of the Froissart theorem.

\item  All those results do nor fix, however, any particular
asymptotic expression for the total cross sections. To obtain
the familiar ``canonical'' restriction of the form $\ln^2 s$,
one should add the hypothesis that in every nonphysical
configuration (even including arbitrary nonphysical angles)
the amplitude cannot grow with energy faster than some finite
power of energy. The Froissart paper~\cite{Fr} ``hides'' this
hypothesis in dispersion relations with a finite number of
subtractions. Note that no physical or mathematical
justifications have been ever suggested for such an asymptotic
hypothesis. Moreover, the observed linearity of Regge
trajectories provides phenomenological arguments against the
power boundedness (more detailed motivation see in Ref.\cite{az}).
In a general case, the upper bound for the total cross section
may grow with energy approximately as the squared logarithm of
the fastest asymptotics of the amplitude in nonphysical
configurations.

\item  The more exact asymptotic expressions for the Legendre
functions, used in Ref.\cite{az}, allowed to strengthen the
original Froissart inequalities~\cite{Fr} for physical
amplitudes (and cross sections). For example, even if the
amplitude  is bounded by a finite power of energy in any
nonphysical configurations, the corresponding total cross
section still {\it cannot} grow as $\ln^2(s/s_0)$ with a fixed
scale $s_0$ (as is usually stated). Instead, the scale $s_0$
itself should grow logarithmically with energy, reducing the
growth rate for the total cross section.

\item Increase of a total cross section faster than the
log-squared energy does not mean violation of unitarity
and is {\it not forbidden} by any general principles,
contrary to a widespread opinion.

\item  It is interesting that neither dispersion relations,
nor any particular properties of interactions were needed
in the analysis of Ref.\cite{az}. The strong interactions,
as an object to apply Froissart restrictions, are marked out
only by the fact of absence of massless hadrons (as
compared, say, to the electrodynamics with its massless
photon).

\end{itemize}

LHC has begun to contribute into the problem of the increasing
total cross sections. The recent analysis of accelerator data
for $pp$ and $p\bar{p}$ scattering~\cite{FMS} assumed the
asymptotic behavior of their total cross sections in the form
$(\ln s)^\alpha $, the exponent $\alpha $ being a free parameter.
The earlier data agree with the ``canonical'' value $\alpha=2 $.
However, addition of the first LHC data~\cite{TOT} appears to
provide small but statistically meaningful excess
$\alpha >2$~\cite{FMS}.

Approach of Ref.\cite{az} enables one to investigate the
high-energy asymptotics not only at a fixed scattering angle
(as in Ref.\cite{Fr}), but also at a fixed momentum transfer.
This allows to study asymptotics of the diffraction peak slope
as well. As appears, if the total cross section increases with
energy, then the diffraction slope should increase at the same
rate or even faster. In the saturation regime, when the total
cross section grows with the maximal possible rate, its ratio to
the slope should stay constant or even decrease~\cite{az}. Such
expectation was in agreement with the pre-LHC accelerator data,
but LHC seems to violate it~\cite{FM}. This means that the
present increase of the total cross sections is not saturated
yet, and when going to even higher energies we may encounter
some unexpected features.

The present work was partly supported by the Russian State grant
RSGSS-65751.2010.2.

\end{document}